\documentclass[aps,pra,twocolumn,showpacs,superscriptaddress,groupedaddress]{revtex4}
\usepackage{graphicx}
\usepackage{dcolumn}
\usepackage{bm}
\usepackage{amssymb}
\usepackage{color}
\usepackage{amsmath}
\usepackage{array}

\usepackage{epstopdf}
\usepackage[latin9]{inputenc}
\usepackage{esint}
\usepackage{bbm}
\usepackage{amsfonts}
\usepackage{mathrsfs}

\begin{document}
\title{Real-time quantum state estimation in circuit QED via  Bayesian approach}

\author{Yang Yang}
\author{Beili Gong}
\author{Wei Cui}
\email{aucuiwei@scut.edu.cn}
\address{School of Automation Science and Engineering, South China University of Technology, Guangzhou 510641, China}

\date{\today}

\begin{abstract}
 Using a circuit QED device, we present a theoretical study of real-time quantum state estimation via quantum Bayesian approach. Suitable conditions under which the Bayesian approach can accurately update the density matrix of the qubit are analyzed. We also consider the correlation between some basic and physically meaningful parameters of the circuit QED and the performance of the Bayesian approach. Our results advance the understanding of quantum Bayesian approach and pave the way to study quantum feedback control and adaptive control.
\end{abstract}

\pacs{42.50.Pq, 42.50.Dv, 85.25.-j}

\maketitle

\section{Introduction}
Quantum state estimation that aims at determining the states of the quantum systems using the measurement records is instrumental to quantum physics experiments and has attracted much attentions \cite{Bisio2009,Doherty2000,Ladd2010}. It can be divided into two categories: quantum state tomography and real-time quantum state estimation. Quantum state tomography completes the task of reconstructing the initial states of the systems \cite{Bisio2009}. There have already been lots of methods applied to achieve efficient quantum state tomography, including maximum likelihood method \cite{Six2016}, linear regression method \cite{Qi2013} and compressive sensing method  \cite{Gross2009,Flammia2012,Kalev2015}. Different with the classical worlds, even though we adopt continuous weak measurements, the quantum states after measurements are still different with the original states due to the back-action of the measurements, the spontaneous decay and dephasing of the systems \cite{Braginsky1992,Wiseman2009}. Thus, the real-time monitoring of the states is widely demanded as well, in which the filter equation \cite{Wiseman2009,Tanaka2012} and the Bayesian approach \cite{Korotkov1999,Korotkov2016} are the two most widely used methods.

Comparing with other methods, the Bayesian approach mainly advantages in its great rapidity and low computational complexity. Many quantum experiments based on the Bayesian approach have been realized \cite{Weber2014,Vijay2012,Gillett2010}. The core content of the quantum Bayesian estimation is about updating the diagonal elements of the qubit\cite{Korotkov1999}, and the methods to update the non-diagonal elements were also raised {recently} \cite{Wang2014,Feng2016,Qin2017,Korotkov2016}. Until now, the Bayesian approach has already been enriched to be an effective method that can be used as a useful tool in quantum information technology and quantum control \cite{Zhang2017}. However, there is no strict proof of its establishment. {Our manuscript} is going to consider the conditions under which the Bayesian approach can real-time precisely monitor the {quantum} states, that is, to analyze the {correlation} between the values of the system parameters and the performance of the Bayesian approach numerically.

This paper is organized as follows. In Sec.~\uppercase\expandafter{\romannumeral2}, {we briefly review the dispersive measurement in circuit QED}. In Sec.~\uppercase\expandafter{\romannumeral3}, we present the quantum  Bayesian approach as well as {examining} its estimation performance.
In Sec.~\uppercase\expandafter{\romannumeral4}, we explore the conditions under which the method can {real-time} precisely track the evolution of the qubit numerically. We plot the error curves of the system parameters and find out all the existing critical points. {We summarize our conclusions in Sec.~\uppercase\expandafter{\romannumeral5}}.

\section{The model}

Superconducting circuit QED has been regarded as a promising candidate for the realization of scalable quantum computing because of its controllability and design flexibility \cite{Blais2004}. Moreover, it is also a great platform to test quantum feedback control and quantum estimation \cite{Xiang2013,Slichter2012,Hatridge2013}. Shortly after being proposed in Ref.~\cite{Blais2004}, it was implemented experimentally \cite{Wallraff2004,Chiorescu2004}. It provides us with several simple high-fidelity readout mechanisms. For example one can realize the continuous weak measurement of the system by performing homodyne measurement on the cavity field.

The circuit-QED system consists of a superconducting qubit and a microwave cavity \cite{Blais2004}. One can describe the qubit, the cavity and their coupling with the Jaynes-Cummings Hamiltonian. In order to realize the measurement of the qubit, the cavity is driven with a tone of amplitude $\varepsilon_d$ and frequency $\omega_d$. The Hamiltonian of the whole system is given by ($\hbar=1$).
\begin{align}
H=&\omega_r a^\dag a+\omega_q\sigma_z/2+g(a\sigma_++a^\dag\sigma_-)\nonumber\\
&+\varepsilon_d(t)e^{-i\omega_dt}a^\dag+\varepsilon^*_d(t)e^{i\omega_dt}a,\label{eq1}
\end{align}
where $\omega_r$ is the cavity frequency, $\omega_q$ is the qubit transition frequency, and $g$ is the coupling strength.  We consider the system in a dispersive regime, $|\Delta|=|\omega_r-\omega_q| \gg |g|$. Under this condition and in the rotating frame, we can use the effective Hamiltonian to describe the system \cite{Gambetta2006}.
\begin{eqnarray}
H_{\rm eff}=\Delta_ra^{\dag}a+{{\tilde\omega_q}\over 2}\sigma_z+\chi a^\dag a\sigma_z+
\varepsilon_d(t)a^\dag+\varepsilon^*_d(t)a,\label{eq2}
\end{eqnarray}
where $\Delta_r=\omega_r-\omega_d$ and $\tilde\omega_q=\omega_q+\chi$ with $\chi=g^2/\Delta$ representing the dispersive coupling strength between the cavity photon number and the qubit.

We use the following quantum stochastic master equation to describe the evolution of the qubit-cavity state $\rho_T(t)$ \cite{Wiseman2009},
\begin{align}
\dot\rho_T(t)=&-{i\over\hbar}[H_{\rm eff},\rho_T(t)]+\kappa D[a]\rho_T(t)+\gamma_1D[\sigma]\rho_T(t)\nonumber\\
&+\gamma_\varphi D[\sigma_z]\rho_T(t)/2+\sqrt{\kappa}H[ae^{-i\phi}]\rho_T(t)\xi(t),\label{eq3}
\end{align}
where $\kappa$, $\gamma_1$ and $\gamma_\varphi$ represent the damping rate of the resonator, qubit decay and dephasing of the qubit, respectively. $\xi(t)$ is a Gaussian white noise coming from the stochastic quantum-jump, and satisfies $E\left[\xi(t)\right]=0$ and $E\left[\xi(t)\xi(t')\right]=\delta(t-t')$. The superoperators $D[c]\rho=c\rho c^\dag-c^\dag c\rho/2-\rho c^\dag c /2$ and
$H[c]\rho=c\rho+\rho c^\dag-{\rm Tr}\left\{\left[c+c^\dag\right]\rho\right\}\rho$.
We simply consider the single quadrature measurement, through which what we actually {measure} is a photocurrent \cite{Gambetta2008}. The photocurrent measured can be expressed as:
\begin{eqnarray}
I_\phi(t)=\sqrt{\kappa\eta}{\left\langle ae^{-i\phi}+a^\dag e^{i\phi}\right\rangle}_t+\xi(t).\label{eq4}
\end{eqnarray}
With the initial state prepared by $\left|e(g)\right\rangle$, we have the coherent state of the cavity field $\left|\alpha_{e(g)}\right\rangle$ whose amplitude evolution can be described with the following equation \cite{Gambetta2006}.
\begin{eqnarray}
\dot\alpha_{e(g)}=-i\varepsilon_d(t)-i(\Delta_r\pm\chi)\alpha_{e(g)}-\kappa\alpha_{e(g)}/2.\label{eq5}
\end{eqnarray}
Ref.~\cite{Gambetta2008} defined $\beta(t)=\alpha_e(t)-\alpha_g(t)=|\beta|e^{-i\theta_\beta}$ to represent the information about the qubit involved in the measurement outcomes.

\begin{figure}[ht]
\centering\includegraphics[width=8.6cm]{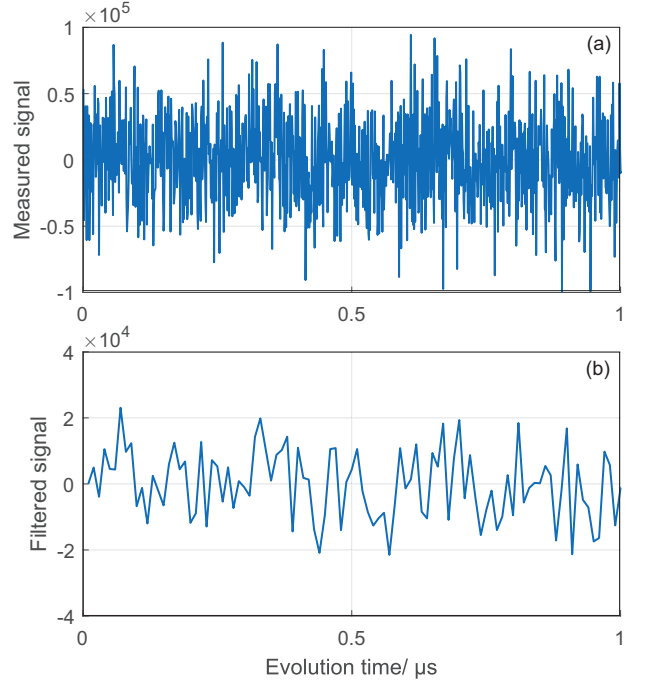}
\caption{(Color online) The measured signal (a) and the filtered signal (b) of a single measurement in circuit QED with $t_m=0.01{\rm\mu s}$, $\omega_q=0$ Hz, $\Delta_r=0$ Hz, $\chi=1$ MHz, $\varepsilon_d=15$ MHz, $\kappa=20$ MHz, $\alpha_e=\alpha_g=0$, $\gamma_1=\gamma_\psi=0$ and $\phi=\pi$. }\label{fig1}
\end{figure}


What one finally want to control is the qubit, so it's necessary to find an equation that gives the evolution of the qubit itself instead of the qubit-cavity system. Through applying a displacement transformation on the Eq.~(\ref{eq3}) and tracing over the cavity state, one can obtain an effective stochastic master equation (SME) for the qubit \cite{Gambetta2008}:
\begin{align}
\dot\rho(t)=&-i{\tilde\omega_q+B(t)\over2}[\sigma_z,\rho(t)]+{\Gamma_d(t)+\gamma_\varphi\over2}D[\sigma_z]\rho(t)\nonumber\\
&+\gamma_1D[\sigma_-]\rho(t)-\sqrt{\Gamma_{ci}(t)}M[\sigma_z]\rho\xi(t)\nonumber\\
&+i{\sqrt{\Gamma_{ab}(t)}\over2}[\sigma_z,\rho]\xi(t),\label{eq6}
\end{align}
where $B(t)=2\chi \rm{Re}[\alpha_g(t)\alpha_e(t)^*]$ describes the generalized ac-Stark shift of the qubit energy due to the dynamic fluctuation of cavity field. The superoperator $M[c]\rho=\left[c-{\rm Tr}(c\rho)\right]\rho/2+\rho\left[c-{\rm Tr}(c\rho)\right]/2$. The effect of the coupling with a cavity on the qubit is transformed to an additional dephasing rate
$\Gamma_d=2\chi \rm{Im}[\alpha_g(t)\alpha_e(t)^*]$. $\Gamma_{ci}(t)=\kappa|\beta(t)|^2\rm{cos}^2(\phi-\theta_\beta)$ and $\Gamma_{ab}(t)=\kappa|\beta(t)|^2\rm{sin}^2(\phi-\theta_\beta)$ respectively represent the measurement efficiency
and the measurement back-action rate for a single measurement \cite{Gambetta2008}. In this reduced representation, the measured photocurrent can also be expressed as
\begin{align}
I(t)&=\sqrt{\Gamma_{ci}}{\left\langle\sigma_z\right\rangle}_t+\xi(t)+\sqrt{\kappa}|\mu(t)|\rm{cos}(\theta_\mu-\phi)
\nonumber\\
&=\bar I(t)+\sqrt{\kappa}|\mu(t)|\rm{cos}(\theta_\mu-\phi),\label{eq7}
\end{align}
where $\mu(t)=\alpha_e(t)+\alpha_g(t)=|\mu|e^{-i\theta_\mu}$. Once we obtain the measurement records during a time interval and want to estimate the system state in the next time with the initial qubit state known, the common  method is the filter equation
\begin{align}
\dot\varrho(t)=&-i{\tilde\omega_q+B(t)\over2}[\sigma_z,\varrho(t)]+{\Gamma_d(t)+\gamma_\varphi\over2}D[\sigma_z]\varrho(t)\nonumber\\
&-\sqrt{\Gamma_{ci}(t)}M[\sigma_z]\varrho(\bar I(t)-\sqrt{\Gamma_{ci}(t)}{\left\langle\sigma_z\right\rangle}_t)\nonumber\\
&+i{\sqrt{\Gamma_{ab}(t)}\over2}[\sigma_z,\varrho](\bar I(t)-\sqrt{\Gamma_{ci}(t)}{\left\langle\sigma_z\right\rangle}_t)~\nonumber\\
&+\gamma_1D[\sigma_-]\rho(t).
\label{eq8}
\end{align}
However, it is time-consuming and computationally complex  to calculate the evolution of the  filter equation,  while  the time consumed to real-time estimation of the quantum states is hoped to be as short as possible in quantum feedback control.  Thus, Korotkov put forward the quantum Bayesian approach \cite{Korotkov1999,Korotkov2016} to deal with this problem.

\begin{figure*}
\centering\includegraphics[width=17.9cm]{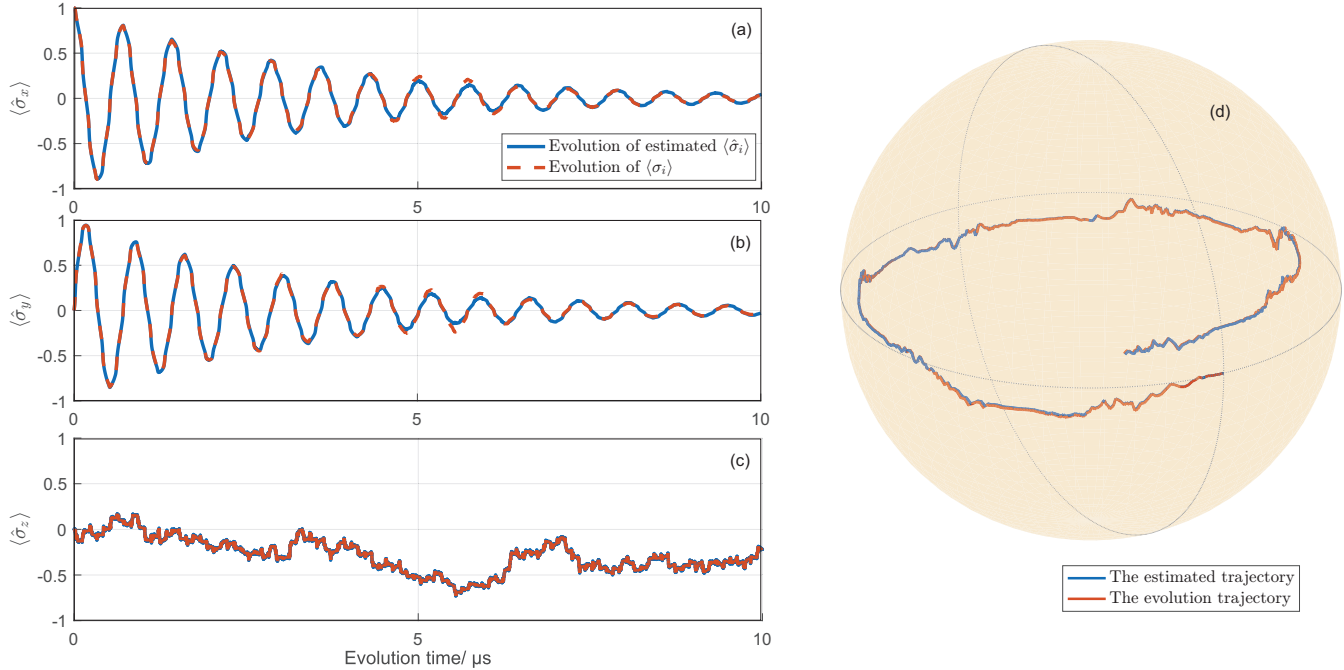}
\caption{(Color online) {Real-time quantum state estimation via quantum Bayesian approach.} {Figs.~2(a)-2(c) detailly depict the estimation of the three Bloch components $\langle\hat\sigma_i\rangle$ with quantum Bayesian approach (solid blue curves) and the original evolution of $\langle\sigma_i\rangle$ (dash red curves) under the standard SME (6). Fig.~2}(d) depicts the trajectories in the Bloch sphere, where the blue {curve} is the estimated trajectory and the red { curve} is the { evolution} one under the standard SME (6).
To make it easier to observe, we only intercept the first $0.75\rm\mu s$ of the trajectory. Here $\eta=0.1$ and the rest parameters are the same as those in Fig.~\ref{fig1}.
}\label{fig2}
\end{figure*}


\section{Bayesian approach for single quadrature measurement}

As we {mentioned} in the previous section, what we actually {measure} in the circuit-QED is a photocurrent that satisfies Eq.~(\ref{eq4}). Due to the noise effect, we apply a low-pass filter process to the photocurrent and average the measured output signal over a time interval $t_m$,
\begin{eqnarray}
{I_m} = \frac{1}
{{{t_m}}}\int_0^{{t_m}} {{I_\phi }(t)dt}.\label{eq9}
\end{eqnarray}
When the initial state is determined, based on Eq.~(\ref{eq5}) we can calculate the evolution of the amplitude of the cavity field. Thus we have the ensemble-average filtered output
\begin{eqnarray}
{{\bar I}_{g(e)}} = \frac{1}{{{t_m}}}\int_0^{{t_m}}{(2\sqrt\kappa)\operatorname{Re}[{\alpha_{g(e)}}(t){e^{-i\phi}}]}dt.
\label{eq14}
\end{eqnarray}
The measurement output can be proved to satisfy a determined Gaussian distribution\cite{Korotkov1999,Wang2014}:
\begin{eqnarray}
{P_{g(e)}}({I_m}) = \frac{1}{{\sqrt {2\pi D} }}\exp [ - {({I_m} - {\bar I_{g(e)}})^2}/2D_I],\label{eq10}
\end{eqnarray}
where $D_I=1/t_m$ characterizes the variance of the distribution. Here we plot the measured signal and filtered signal of an individual measurement in Fig.\ref{fig1}.

As the diagonal elements of the qubit density matrix describe the probability distribution of the system state, they can be regarded as the priori knowledge. With the determined distribution Eq.~(\ref{eq10}) and the measurement output, the classical Bayesian formula is proposed to update the diagonal elements of the quantum state.
\begin{align}
&{\hat\rho _{gg}}({t_m}) = {\rho _{gg}}(0){P_g}({I_m})/N,\nonumber\\
&{\hat\rho _{ee}}({t_m}) = {\rho _{ee}}(0){P_e}({I_m})/N,\nonumber\\
&N = {\rho _{gg}}(0){P_g}({I_m}) + {\rho _{ee}}(0){P_e}({I_m}).\label{eq11}
\end{align}
If the qubit is a pure state, one can approximate $\rho_{ge}(t_m)$ as \cite{Korotkov1999}
\begin{eqnarray}
{\tilde\rho _{ge}}({t_m}) = {\rho_{ge}}(0){e^{-i{{\tilde\omega}_q}{t_m}}}\sqrt{{P_g}({I_m}){P_e}({I_m})}/N.\label{eq12}
\end{eqnarray}
However this result was rather imprecise and quite different with the filter equation (\ref{eq8}) due to the purity degradation and stochastic fluctuations of the systems. Thus, in order to further correct the estimation of the non-diagonal elements, Ref.~\cite{Korotkov2016} detailly analyzed the measurement-induced backaction and  found that besides the diagonal part the phase backaction, the dephasing due to nonideality of the measurement and ac Stark {shift} of the qubit frequency may play more significant role in producing evolution of the non-diagonal elements.
 {Based on the stochastic master equation (\ref{eq6}),  Refs.~\cite{Wang2014,Feng2016} further considered the cavity-field-fluctuation effects on the qubit and
  proposed a different method to accurately estimate the non-diagonal matrix elements. } The updated quantum Bayesian approach can be summarized as follows
\begin{align}
&{\hat\rho_{ge}}({t_m}) = {{\tilde\rho}_{ge}}({t_m})De(t_m)\exp\{- i[{\Phi_1}({t_m}) + {\Phi_2}({t_m})]\},\nonumber\\
&De(t_m)=\left|\langle{{\alpha_e}({t_m})}|{{\alpha_g}({t_m})}\rangle\right|,\nonumber\\
&{\Phi_1}({t_m}) = \int_0^{{t_m}}{B(t)dt},\nonumber\\
&{\Phi_2}({t_m}) = -\int_0^{{t_m}}{\sqrt{{\Gamma_{ba}}(t)}\bar I_\phi(t)dt},\nonumber\\
&\bar I_\phi(t) = I_\phi(t)-\sqrt{\kappa}|\mu|\cos(\theta_\mu-\phi),\label{eq13}
\end{align}
where $De$ is the purity degradation factor, $\Phi_1$ and $\Phi_2$ are the two additional phase factors.

With Eqs.~(\ref{eq11}) and Eqs.~(\ref{eq13}) we can {real-time estimate the quantum state}. {In Fig.~\ref{fig2}, we plot the the estimation of the three Bloch components $\langle\hat\sigma_i\rangle$ with Bayesian approach and the evolution of $\langle\sigma_i\rangle$. Moreover the estimated trajectory and the evolution trajectory are depicted in Fig.~2(d), from which it is clear that the Bayesian approach can real-time accurately estimate the quantum state of the circuit QED system. To make it easier to observe, we only intercept the first 0.75$\mu s$ of the trajectories. Here $\eta=0.1$ and the rest parameters are the same as those in Fig.~\ref{fig1}.}

\section{Suitable conditions of the Bayesian approach}

Although the quantum Bayesian approach provides us with great rapidity and low computational complexity, there is no theoretical proof about its establishment, and it still remains an assumption. So exploring the suitable conditions of the quantum Bayesian approach might be a critical issue.
For simplicity, we consider some basic and physical meaningful parameters involved in the experiment stated above and concentrate on finding out the ranges of these parameters in which the Bayesian approach can work properly. The reduced master equation Eq.~(\ref{eq6}) is regarded as the standard evolution of the qubit. It should be noted that the system works in the dispersive regime, $|\Delta|=|\omega_r-\omega_q| \gg |g|$. This regime limits $\chi$ to be much smaller than $|\Delta|$ (usually approximately equal to $\omega_q$).

\begin{figure}
\centering\includegraphics[width=8.6cm]{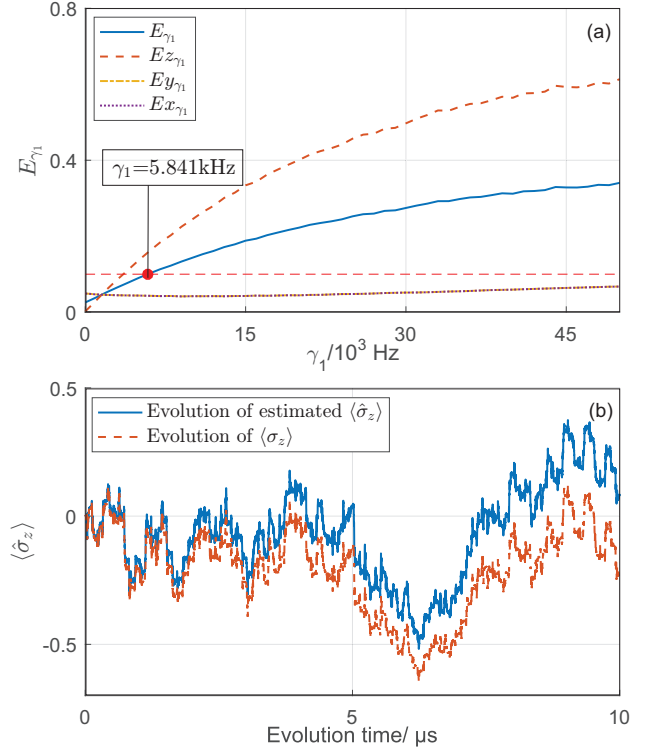}
\caption{(Color online) Evolution of the error functions $E_{\gamma_1}$, $Ez_{\gamma_1}$, $Ey_{\gamma_1}$ and $Ex_{\gamma_1}$ for a total time of 10$\mu s$ and with 5000 trajectories, as  functions of the parameter $\gamma_1$. In Fig.~3(a) the blue solid curve, the red dashed curve, the yellow pecked curve and the purple dotted curve represent $E_{\gamma_1}$ $Ez_{\gamma_1}$, $Ey_{\gamma_1}$ and $Ex_{\gamma_1}$, respectively. The red filled dot is the critical point of $\gamma_1$ at which $E_{\gamma_1}$ reaches 0.1.  For clarity, the time evolution of the estimated $\langle\hat{\sigma}_z\rangle$ (solid blue curve) and the evolution of $\langle\sigma_z\rangle$ (dash red curve) at this point are plotted in Fig.~3(b). }\label{fig3}
\end{figure}


\begin{figure*}
\centering\includegraphics[width=17.9cm]{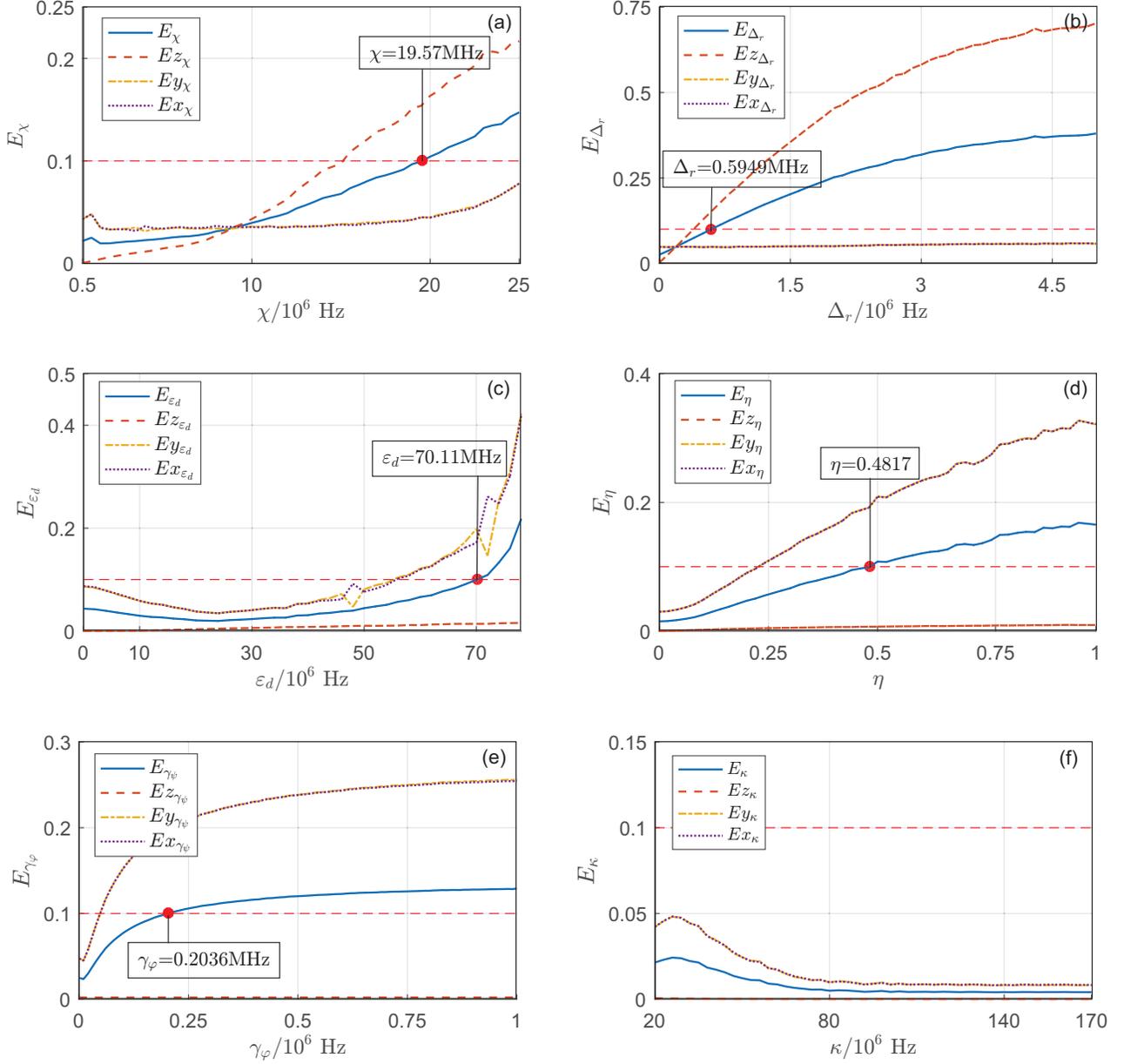}
\caption{(Color online)  Evolutions of the error functions $E_{a}$, $Ex_{a}$, $Ey_{a}$ and $Ez_{a}$ for a total time of 10$\mu s$ and with 5000 trajectories, as  functions of the parameters $\chi$, $\Delta_r$, $\varepsilon_d$, $\eta$, $\gamma_\varphi$ and $\kappa$ in Figs.~4(a)-4(f), respectively. The blue solid curves, the red dashed lines, the yellow pecked curves and the purple dotted curves represent $E_{a}$ $Ez_{a}$, $Ey_{a}$ and $Ex_{a}$ respectively. The red filled dot in each subfigure represents  at which the overall error $E_{a}$ reaches 0.1.
}\label{fig4}
\end{figure*}


The basic and physical meaningful parameters in the circuit-QED system we considered are $\gamma_1$, $\gamma_\varphi$, $\chi$, $\varepsilon_d$, $\Delta_r$, $\kappa$ and $\eta$. For a specific parameter $a$, we characterize the overall performance of the Bayesian approach by defining an error $E_a$, and the performances of the three Bloch components ($\langle\sigma_x\rangle$, $\langle\sigma_y\rangle$, $\langle\sigma_z\rangle$ ) estimation are defined as $Ex_a$, $Ey_a$, and $Ez_a$, respectively.  Due to the dissipation property of the system, the ensemble evolution of the master equation (\ref{eq6}) would tend to be a steady state $\rho^{\ast}$, or ($\langle\sigma_x^{\ast}\rangle$, $\langle\sigma_y^{\ast}\rangle$, $\langle\sigma_z^{\ast}\rangle$). To make the performance analysis more accurate, we consider the error functions $Ex_a$, $Ey_a$, and $Ez_a$ are defined in an unstable time interval
 \begin{eqnarray}
 S_i&=&\bigg\{t:\bigg|\langle\sigma_i(t)\rangle-\langle\sigma_i^{\ast}(t)\rangle\bigg|>\epsilon_1 \bigg\}\nonumber\\
 &&\cup\bigg\{t:\bigg|\langle\hat{\sigma}_i(t)\rangle-\langle\sigma_i^{\ast}(t)\rangle\bigg|>\epsilon_1 \bigg\},~~~~~~i=x,~y,~z,\nonumber\\
 \end{eqnarray}
where $\epsilon_1$ is a  steady-state precision.  Throughout this manuscript we set  $\epsilon_1=1\%$. For a given parameter $a$, the error functions $Ex_a$, $Ey_a$, and $Ez_a$ are defined as
\begin{eqnarray}
Ex_a=\int_{S_x}\frac{[\langle\sigma_x(t)\rangle-\langle\hat{\sigma}_x(t)\rangle]^2}{\parallel S_x\parallel}dt,\nonumber\\
Ey_a=\int_{S_y}\frac{[\langle\sigma_y(t)\rangle-\langle\hat{\sigma}_y(t)\rangle]^2}{\parallel S_y\parallel}dt,\\
Ez_a=\int_{S_z}\frac{[\langle\sigma_z(t)\rangle-\langle\hat{\sigma}_z(t)\rangle]^2}{\parallel S_z\parallel}dt. \nonumber
\end{eqnarray}
Thus, we can calculate the overall error for a specific parameter $a$
\begin{eqnarray}
E_a=\alpha_x{Ex}_a+\alpha_y{Ey}_a+\alpha_z{Ez}_a,
\end{eqnarray}
where $\alpha_x$, $\alpha_y$ and $\alpha_z$ are the weighting factors used to achieve a balance between the three Bloch components. It is clear that the diagonal elements play a more important role in the quantum Bayesian approach. Thus,  throughout this manuscript we set $\alpha_x=\alpha_y=0.25$, and $\alpha_z=0.5$.

\renewcommand\arraystretch{2}
\begin{table*}
\caption{The suitable conditions of the system parameters ($E_a\le0.1$)}
\begin{tabular}{|p{2.7cm}<{\centering}|p{2cm}<{\centering}|p{2cm}<{\centering}|p{2cm}<{\centering}|p{2cm}<{\centering}|p{2cm}<{\centering}|p{2cm}<{\centering}|p{1.9cm}<{\centering}|}
\hline
Parameter & $\chi$ & $\Delta_r$ & $\varepsilon_d$ & $\eta$ & $\gamma_1$ & $\gamma_\varphi$ & $\kappa$\\
\hline
Suitable conditions & $\le19.57$MHz & $\le0.5949$MHz & $\le70.11$MHz & $\le0.4817$ & $\le5.841$kHz & $\le0.2036$MHz & -\\
\hline
\end{tabular}
\label{table1}
\end{table*}
\begin{table*}
\caption{The suitable conditions of the system parameters ($E_a\le0.05$)}
\begin{tabular}{|p{2.7cm}<{\centering}|p{2cm}<{\centering}|p{2cm}<{\centering}|p{2cm}<{\centering}|p{2cm}<{\centering}|p{2cm}<{\centering}|p{2cm}<{\centering}|p{1.9cm}<{\centering}|}
\hline
Parameter & $\chi$ & $\Delta_r$ & $\varepsilon_d$ & $\eta$ & $\gamma_1$ & $\gamma_\varphi$ & $\kappa$\\
\hline
Suitable conditions & $\le12.1$MHz & $\le0.1959$MHz & $\le53.4$MHz & $\le0.2372$ & $\le1.821$kHz & $\le0.0467$MHz & -\\
\hline
\end{tabular}
\label{table2}
\end{table*}
\renewcommand\arraystretch{0.5}

In order to maintain the estimation accuracy one can set an error upper-bound $\epsilon_2$, and make the overall error $E_a\le \epsilon_2$.   Once $E_a$ exceeds $\epsilon_2$, the Bayesian approach is considered unable to work properly. Without loss of generality, we set $\epsilon_2=0.1$ in our numerical experiments. Moreover, we find that $E_a$ exceeds $\epsilon_2$ only when certain parameter becomes too large, thus we define the point at which $E_a$ reaches $\epsilon_2$ as the critical point. When certain parameter exceeds its critical point, the Bayesian approach is regarded as inapplicable.
In the following we numerically find the critical value of a specific parameter $a$, below or above which the Bayesian approach works well. We keep the other parameter constant and make the parameter $a$ change within a certain range. Based on the ensemble evolution of the system state, we calculate the evolution of the error function $E_a$ with respect to the parameter $a$. In this way, we can obtain the error curves of these parameters, which describe the relationships between the values of the parameters and the performance of the Bayesian approach.

In Fig.~\ref{fig3}, we take $\gamma_1$ for example.
 The error curves of $\gamma_1$ for a total time of 10$\mu$s and with $5000$ single evolutions are illustrated in Fig.~\ref{fig3}(a).
 The blue solid curve, the red dashed curve, the yellow pecked curve and the purple dotted curve represent $E_{\gamma_1}$, $Ez_{\gamma_1}$, $Ey_{\gamma_1}$ and $Ex_{\gamma_1}$, respectively.
 The red filled dot is the critical point of $\gamma_1$ at which $E_{\gamma_1}$ reaches $\epsilon_2$.
 It is clear that the increasing of $\gamma_1$ mainly results in increasing of $Ez_{\gamma_1}$ and the critical point of $\gamma_1$ is around $5.841$ kHz.
  For clarity, the time evolution of the estimated $\langle\hat{\sigma}_z\rangle$ (solid blue curve) and the evolution of the original $\langle\sigma_z\rangle$ (dash red curve) at the critical point are plotted in Fig.~3(b) from which one can see that the Bayesian approach doesn't work well under such a parameter setting. This result demonstrates
the effectiveness of the proposed error definition.

The other parameters in the circuit-QED system we considered are $\chi$, $\Delta_r$, $\varepsilon_d$,  $\eta$, $\gamma_\varphi$, and $\kappa$.
 We find that the values of all parameters except $\kappa$ have the ability to result in rejecting of the Bayesian approach.  The phenomenon of the errors evolution with respect to $\kappa$  is plotted in Fig.~4(f). This is reasonable because $\kappa$ represents the damping rate of the  resonator and only has an effect on the photocurrent Eq.~(\ref{eq4}).
  The rest error curves of the parameters considered are also plotted in Fig.~\ref{fig4}, from which the critical points can be easily found to be around 19.57MHz, 0.5949MHz, 70.11MHz, 0.4817 and 0.2036MHz, corresponding to $\chi$, $\Delta_r$, $\varepsilon_d$, $\eta$ and $\gamma_\varphi$, respectively. { All of these five parameters have significant effects on the evolution of the error, where $\chi$ and $\Delta_r$ mainly affect the estimation of the diagonal elements of the density matrix. On the contrary, $\varepsilon_d$, $\eta$ and $\gamma_\varphi$ mainly affect the estimation of the non-diagonal elements. For convenience of reference, we conclude  the suitable conditions that allow the overall error $E_a\le0.1$ in Tab.~\ref{table1}.
  The definition of the error $E_a$ as well as the values of the weighting factors can be adjusted according to the specific precision requirement.  In Tab.~\ref{table2} we display the suitable conditions of various systems parameters
  that allows the overall  error $E_a\le0.05$. There is no doubt that the selection becomes much narrower.
  As a preliminary work, this paper only considers the situation under the constraint of a single parameter, while the research about the collaborative effects of the multi-parameters might be of interest for further studies.}

\section{Conclusion}

In conclusion, this paper theoretically {studies} the Bayesian approach which {completes} the task of real-time quantum state estimation in circuit QED. The main content of our work is to analyse the suitable conditions under which the Bayesian approach can real-time accurately estimate the quantum state. In detail we explore the correlation between some basic and physically meaningful parameters of the circuit QED and the performance of the Bayesian approach. Our work helps to gain a deeper understanding of the quantum Bayesian approach and promotes the application of the quantum Bayesian approach in quantum control. {Moreover, through the analysis of the correlation between the system parameters and the performance of the method, the roles that these parameters play in the system can be understood more vividly.}

\begin{acknowledgements}
This work is supported by National Natural Science Foundation of China under Grant 11404113, and the Guangzhou Key Laboratory of Brain Computer Interaction and Applications under Grant 201509010006.
\end{acknowledgements}

\label{sec:TeXbooks}

\end{document}